\documentstyle[twocolumn,aps,prl,amssymb,psfig,floats]{revtex}
\def\ggt{{\gamma\leftrightarrow\tilde\gamma}}
\def\gtg{{\gamma\tilde\gamma}}
\def\n{\textsf{n}}
\begin{document}
\title{The Photino Induced Distortion of the CMBR Blackbody Spectrum}
\author{M.V. Medvedev \cite{mm}}
\address{Harvard-Smithsonian Center for Astrophysics, 60 Garden St., 
Cambridge, MA 02138}
\maketitle
\draft
\begin{abstract}
It is shown that photon--photino inter-conversions (if exist) may result in a 
detectable CMBR spectrum distortion which amplitude depends on photino 
properties, such as its mass. An upper bound on the distortion parameter 
determined from the recent COBE-FIRAS data, hence, sets 
a lower bound on the photino mass, $m_{\tilde\gamma}\gtrsim300$~eV.
\end{abstract}

The elegant and generally accepted theory of super-symmetry in 
particle physics (for astrophysical implications, see, e.g., \cite{SUSY}) 
which unifies bosons and fermions remains
neither confirmed nor rejected experimentally so far. This theory predicts
the existence of new particles, the super-symmetric partners 
for all known particles, neither of which has been detected yet.
In this Letter, we show that, under particular conditions, the photino
$\tilde\gamma$ (the $\frac{1}{2}$-spin fermionic super-symmetric 
partner of the photon $\gamma$), if exists, may result in a measurable 
distortion of
the blackbody Cosmic Microwave Background Radiation (CMBR) energy 
spectrum. The magnitude of the deviation from the Planckian spectrum is 
directly related to the value of the photino mass, $m_{\tilde\gamma}\not=0$.
We then use recent results \cite{COBEresults} from the COsmic Background 
Explorer (COBE) mission to put some constraints on the value of 
$m_{\tilde\gamma}$. Recent evidence for the time variability of physical 
constants (e.g., the fine structure constant) \cite{varyingconst} makes
the measurement of the `photino induced' CMBR spectrum distortion to be of
great importance. Indeed, the CMBR provides information on physical laws
existed in the early Universe (at redshifts $z\sim 10^3-10^8$) which
cannot be obtained by means of other astronomical observations \cite{c1}.
Thus, such an effect, if detected, would have profound implications for 
our understanding of fundamental physics.

Photons during the radiation dominated era ($10^3\lesssim z\lesssim10^9$)
are in thermal and ionization equilibrium with matter. However, the 
process of photon energy equilibration throughout the spectrum  is
quite slow, because it proceeds via the (inverse) Compton scattering
with electrons \cite{Syun-Zeld}, which number density is small with
respect to the number density of photons, $\n_e/\n_\gamma\sim10^{-9}$.
In fact, the rate of Comptonization of photons becomes comparable to 
the Universe expansion rate at redshift 
$z_c\sim{2\times10^4\Omega_0^{-1/2}}\sim10^5$ (where 
$\Omega_0=\rho_0/\rho_{crit}$ is the ratio of present density of matter 
in the Universe to the critical density). Thus, the characteristic time, 
$\tau_{eq}$, required to establish the (spectral) equilibrium in the 
photon gas diverges as $z$ approaches $z_c$.

On the other hand, in the super-symmetry theory, the photon and photino
are different ``spin states'' (with ${s=1}$ and ${s=\frac{1}{2}}$,
respectively) of a particle and have different
interchange properties. The photon is known to be a massless particle, and
the photino is believed to be  massive. Thus, if a corresponding mixing 
angle $\theta$ of these two `states' is not identically zero, one may
expect the $\ggt$ oscillations to occur, with the probability
$P_{\gamma\to\tilde\gamma}^{osc}(t)=\frac{1}{2}\sin^2{2\theta}
\left[1-\cos\left((E_{\tilde\gamma}-E_\gamma)\hbar^{-1}t\right)\right]$,
where the energy difference is 
$E_{\tilde\gamma}-E_\gamma\sim m_{\tilde\gamma}c^2$.
Such a simple picture, however, is not correct, because the oscillations 
of a single particle are forbidden due to angular momentum and
R-parity conservation. Thus, the $\ggt$ inter-conversions are allowed 
to occur only in reactions (collisions) with other particles. (We postpone the
discussion of probable reaction channels intill the end of the Letter.)
Consequently, the average rate $\nu_\gtg$ of
$\ggt$ inter-conversions is, thus, the average collision frequency,
$\langle\sigma_{\tilde\gamma}v\rangle \n$, times the probability
of a $\ggt$ oscillation during collision time, 
$P_{\gamma\to\tilde\gamma}^{osc}(\Delta t_{coll})$,
(note, $P_{\gamma\to\tilde\gamma}^{osc}=
P_{\tilde\gamma\to\gamma}^{osc}$, by symmetry). It should be 
emphasized that $\ggt$ inter-conversions are incoherent in this regime,
i.e., the quantum coherence of photon-photino oscillations is
completely lost at times greater than $\nu^{-1}_\gtg$ due to
classical, intrinsically chaotic motion of interacting particles.
Photinos are, hence, in (at least, marginal) equilibrium with photons.
The equilibrium fraction of $\tilde\gamma$'s in a system is
proportional to the forward--to--reverse transition rate ratio and,
hence, is independent of the mixing angle. The rate with 
which equilibrium is established is, however, affected by $\theta$.
We now introduce the total characteristic time of a $\ggt$ transition
as follows, $\tau_\gtg\simeq\tau_{\gamma\to\tilde\gamma}+
\tau_{\tilde\gamma\to\gamma}$. The numbers of photons and photinos
in the system, thus, fluctuate on the time-scale $\tau_\gtg$,
while their total number is preserved.

Let us consider two limiting cases. First, the spectral equilibration
time is much smaller than the typical $\ggt$ fluctuation time,
$\tau_{eq}\ll\tau_\gtg$. This is a general case, because
super-symmetric particles 
are believed to be weakly interacting (otherwise they would be detected).
In this case, the number of photons in the system fluctuates adiabatically 
slowly and CMBR photons have enough time to re-arrange throughout the
spectrum. Hence, the Planckian (or Bose-Einstein \cite{Syun-Zeld})
spectrum is maintained. Second, if the spectral thermalization process
is (externally) inhibited, e.g., by the Universe expansion, the
opposite case may realize. Then
\begin{mathletters}
\label{conditions}
\begin{eqnarray}
& &\tau_\gtg\ll\tau_{eq}\ , \label{condition1}\\
& &\tau_\gtg\lesssim t_{expan}\sim(3H)^{-1}\ , \label{condition2}
\end{eqnarray}
\end{mathletters}
(where $H$ is the instantaneous Hubble constant) and $\ggt$ transitions
are faster than the photon transit time though the spectrum. 
Thus, different, `mean-field' equilibrium, which is, in general, a
weighted sum over all realizations, is created;
rather than the `instantaneous' Planckian equilibrium. [Note, inequality 
(\ref{condition2}) ensures that $\ggt$ inter-conversions are fast enough,
compared to the expansion rate]. The partition function 
and occupation numbers for a gas of particles with fluctuating quantum 
statistics were calculated elsewhere \cite{ambiguous}.

Since the CMBR spectrum is very close to Planckian with temperature
$T_0\simeq2.7$~K, we look for a small deviation only. Then, two free
parameters, $p_f$ and $p_b$ (the boson and fermion probabilities, see
Ref.\ \cite{ambiguous} for details), become $p_b=1-p_f$ and $p_f\ll1$.
Thus, we may expand the occupation number $n$ in terms of $p_f$.
From Eqs.\ (11,12) of Ref.\ \cite{ambiguous,c2}, we straightforwardly
obtain the {\em effective} bosonic and fermionic contributions
(${n_b+n_f=n}$):
\begin{mathletters}
\label{nb+nf}
\begin{eqnarray}
& &n_\gamma\equiv n_b=\frac{1}{e^x-1}
\left(1+\sqrt{p_f}\sqrt{\frac{e^x}{e^x-1}}\right) \ ,
\label{nb}\\
& &n_{\tilde\gamma}\equiv n_f=\sqrt{p_f}\frac{1}{\sqrt{e^x(e^x-1)}} \ ,
\label{nf}
\end{eqnarray}
\end{mathletters}
where $x=\epsilon/kT$ is the dimensionless energy. Thus, the deviation of 
the photon distribution from the Planckian one is
\begin{equation}
\delta n_\gamma=n_\gamma-n_{Pl}=\sqrt{p_f}\sqrt{\frac{e^x}{(e^x-1)^3}} \ .
\label{delta-n}
\end{equation}
The parameter $p_f$ is related to the photino mass, $m_{\tilde\gamma}$,
as follows. Let us notice that condition (\ref{condition2}) states
that photinos are in equilibrium with photons: 
$\partial_t\n_{\tilde\gamma}=
P_{b\to f}\n_\gamma-P_{f\to b}\n_{\tilde\gamma}=0$. 
The probability to change a `state' from bose to fermi, $P_{b\to f}$, is 
complimentary to that to stay in a bose `state', $p_b$.
Hence $P_{b\to f}=1-p_b\equiv p_f$, and we obtain:
\begin{equation}
p_f\simeq\frac{p_f}{p_b}=\frac{\n_{\tilde\gamma}}{\n_\gamma} \ .
\end{equation}
The number densities of photons and photinos are calculated integrating
Eqs.\ (\ref{nb+nf}) over energies. Photons are relativistic 
(${\epsilon=pc}$) and the degeneracy $g_\gamma=2$ (due to two
polarizations); photinos are assumed non-relativistic 
($m_{\tilde\gamma}c^2\gg kT_c$) and $g_{\tilde\gamma}=(2s+1)=2$. 
Finally, we obtain:
\begin{equation}
\sqrt{p_f}\simeq\frac{1}{\zeta(3)}\sqrt{\frac{\pi}{8}}
\left(\frac{m_{\tilde\gamma}c^2}{kT_c}\right)^{3/2}
e^{-m_{\tilde\gamma}c^2/kT_c} \ ,
\label{pf}
\end{equation}
where $\zeta$ is the Riemann $\zeta$-function, $\zeta(3)\simeq1.2$
and $T_c$ is the temperature at redshift $z_c$, 
$T_c\simeq T_0z_c\sim{2.7\times10^5}$~K,
to satisfy condition (\ref{condition1}). 

The CMBR spectrum has been measured with great accuracy by the FIRAS
(Far-InfraRed Absolute Spectrophotometer) on board of the COBE satellite.
For our analysis, we use the CMBR residuals (after subtraction of the
Planck blackbody and galactic emission spectra) of both monopole 
and dipole components of Ref.\ \cite{COBEresults} (referred to as the
`96-monopole' and `96-dipole' data sets, respectively) and, 
for comparison, some older data of Ref.\ \cite{94mono} 
(`94-monopole' data set) and Ref.\ \cite{94dipole} (`94-dipole' data set).
Note, if a dipole component is associated with the Earth motion only
(with respect to the CMBR), both monopole and dipole spectra should
be identical. In general, a dipole spectrum is less sensitive to
systematic errors (e.g., absolute calibration), but statistical errors
may be higher, because a dipole component is of much smaller amplitude 
than monopole.

\begin{figure}
\psfig{file=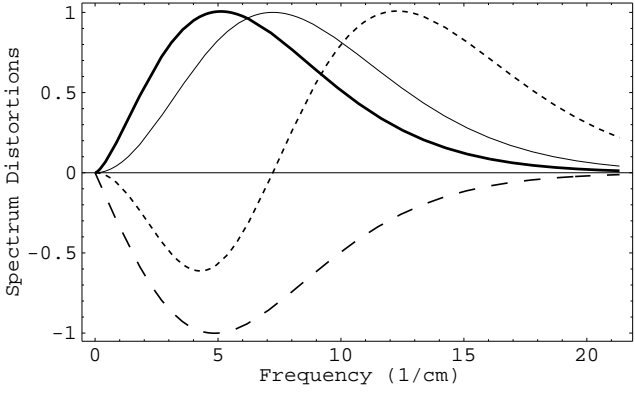,width=3.3in}
\caption{The normalized spectrum distortions 
$(\partial S_\alpha/\partial\alpha)/
\left|(\partial S_\alpha/\partial\alpha)\right|_{max}$ for the
$\sqrt{p_f}$-model (bold solid line), $\mu$-model (dashed line), 
$y$-model (dotted line), and the effect of a temperature shift 
$(\partial B_\nu/\partial T)
/\left|(\partial B_\nu/\partial T)\right|_{max}$ (solid line) 
vs. frequency $\nu$.}
\label{fig}
\end{figure}

Since deviations from the Planck spectrum are small, a linear fit can
be performed:
\begin{equation}
I_{residual}(\nu)=\Delta T\frac{\partial B_\nu}{\partial T} +
\sum_{models (\alpha)}
\alpha\frac{\partial S_\alpha(\nu)}{\partial\alpha}\ .
\label{fit}
\end{equation}
Here $B_\nu$ is the Planck blackbody spectrum, $S_\alpha$ represents
a spectral model, and $\alpha$ is a distortion parameter
which quantify the deviation from $B_\nu(T_0)$. It is important to
have the first term since $\partial B_\nu/\partial T$ correlates
with other models $\partial S_\alpha/\partial\alpha$.
We consider three spectral models. First, the `photino induced'
distortion (referred to as the $\sqrt{p_f}$-model):
\begin{equation}
\frac{\partial S_{\sqrt{p_f}}}{\partial \sqrt{p_f}}
=2hc^2\nu^3\sqrt{\frac{e^{x_0}}{\left(e^{x_0}-1\right)^3} }\ ,
\end{equation}
which readily follows from Eq.\ (\ref{delta-n}), where $x_0=hc\nu/kT_0$
and $\nu$ is measured in cm$^{-1}$. The distortion parameter  
is $\sqrt{p_f}$. Second, the Bose-Einstein quasi-equilibrium photon
distribution \cite{Syun-Zeld} with dimensionless chemical potential
$\mu$ ($\mu$-model). Third, the Comptonized spectrum
characterized by the Kompaneets \cite{Kompaneets} parameter $y$
($y$-model). (For details regarding the last two models, see, e.g., 
Ref.\ \cite{COBEresults}.) Note that all three distortion parameters,
$\sqrt{p_f}$, $\mu$, and $y$, must be {\em positive} by physical meaning.
The normalized spectrum distortions and the effect of a temperature shift
$\partial B_\nu/\partial T$ are shown in Fig.\ \ref{fig}.
Note very strong similarity of the $\sqrt{p_f}$- and $\mu$-models,
which precludes their simultaneous fit. However, preference
to one of the models may be given, due to their {\em opposite}
contributions to the CMBR spectrum. 

Fits (\ref{fit}) have been performed
with the help of {\sc Mathematica~3.01} standard package for
{\em Linear Regression} analysis, which uses the weighted least square
$[\sum_\nu(\Delta_\nu/\sigma_\nu)^2/\sum_\nu(1/\sigma_\nu)^2]$ 
method, where $\Delta_\nu$ is the difference of the model and the data at 
frequency $\nu$, and $\sigma_\nu$ is the uncertainty. Test fits of the
$y$-model and the $\mu$-model separately to the both 96- and 
94-monopole  spectra yield values of $y$ and $\mu$ consistent with those of 
Refs.\ \cite{COBEresults,94mono}. In the dipole fits, $y$ and $\mu$
tend to negative (unphysical) values, so do their simultaneous fits,
i.e., the ${(\mu+y)}$-model, for both monopole and dipole components.
All the fits, which include the both $\mu$- and $\sqrt{p_f}$-models, yield
large statistical errors resulting from very strong correlation
of these models, so they have been rejected. Thus, there are only 
two statistically reliable models, namely the $\sqrt{p_f}$-model and the
${(\sqrt{p_f}+y)}$-model. The results with $1\sigma$ uncertainties
(95\% confidence level) are presented in Table\ \ref{table}.
\begin{table}
\caption{Results of fits for $\sqrt{p_f}$- and ${(\sqrt{p_f}+y)}$-models.}
\begin{tabular}{lr@{$\pm$}lr@{$\pm$}lr@{$\pm$}l}
\ & 
\multicolumn{2}{c}{$\sqrt{p_f}$-model} & 
\multicolumn{4}{c}{${(\sqrt{p_f}+y)}$-model} \\[.5ex]
\cline{2-3}\cline{4-7}\\[-2.5ex]
Data set & 
\multicolumn{2}{c}{$\sqrt{p_f}\times10^5$} &
\multicolumn{2}{c}{$\sqrt{p_f}\times10^5$} & 
\multicolumn{2}{c}{$y\times10^6$} \\ 
\hline 
96-monopole &  1&4 &  3&8  &  2&6  \\
96-dipole   & \ \ \ 13&2 & \ \ 26&4  & 11&3  \\
94-monopole & 10&9 & 50&14 & 35&11 \\
94-dipole   &  7&5 &  3&9  & $-$4&6  \\
\end{tabular}
\label{table}
\end{table}
There is an obvious inconsistency between the 96-monopole and 96-dipole
fits for both models. Since the results for the 96-dipole and 94-monopole
spectra calculated in the $\sqrt{p_f}$-model are also inconsistent with
those in the ${(\sqrt{p_f}+y)}$-model (even within $3\sigma$ uncertainty),
we consider them as statistically unreliable. What causes such
anomalously large correlations of residuals (especially in the
96-dipole data set) is not known. This issue requires further
investigation. (One of possible `candidates' is residual 
galactic contamination, which is strongly coupled to the $y$-model.)
Other fits, i.e., for the 96-monopole and 94-dipole data, with 
both models yield results which are consistent with each other and
include the {\em `null'} result (${\sqrt{p_f}=y=0}$) within 
$\sim1.4\sigma$ uncertainty.
Weak trend toward positive values may, however, be discerned.
Using the values with smaller statistical uncertainties, we set the 
following conservative upper limit on the parameter $\sqrt{p_f}$:
\begin{equation}
\sqrt{p_f}\lesssim10^{-4}\ .
\label{bound-pf}
\end{equation}
From Eq.\ (\ref{pf}), a lower bound on the photino mass is
\begin{equation}
m_{\tilde\gamma}\gtrsim300\textrm{~eV}/c^2\ ,
\label{bound-m}
\end{equation}
which is much lower \cite{c1} than generally accepted:
$m_{\tilde\gamma}\sim0.1-10\textrm{~GeV}/c^2$.
Note, however, that the {\em `null'} result cannot yield absolutely
reliable lower bound on $m_{\tilde\gamma}$, 
because such a result can also be due to breaking of condition 
(\ref{condition2}), which accounts for $\ggt$ kinetics.

Now we check the consistency of our result
with conditions (\ref{conditions}). In fact, we should check 
(\ref{condition2}) only, because (\ref{condition1}) is satisfied at 
$z\sim z_c$ by definition. 
Since binary collisions are most frequent, the leading processes are:
(i) the binary photino collisions, 
${\gamma\gamma\rightleftarrows\tilde\gamma\tilde\gamma}$;
(ii) the R-parity `exchange',
${\gamma\tilde x\rightleftarrows\tilde\gamma x}$, 
where $x$ is any particle (possibly an electron, $e$, as the most
abundant) and `tilde' denotes its super-symmetric partner; and (iii) the
decay-type processes, ${\tilde\gamma\rightleftarrows\tilde x\gamma}$,
where $\tilde x$ may be an axino, $\tilde a$ (the super-symmetric partner 
of the axion), or another hypothetical particle. We estimate cross sections of 
these processes and show that only the last reaction cannot be completely
ruled out. Of course, the reaction channels that could influence the 
$\ggt$ transitions are not limited to those written above and may include,
for instance, `chain reactions', 
$\gamma\tilde x_1\to\tilde\gamma x_2$, 
$\tilde\gamma x_3\to\gamma\tilde x_4$, etc.
Since complete study of $\ggt$ reactions
is beyond the scope of this Letter, we restrict ourselves considering
the transition processes mentioned above. 
Since the results below are illustrative, we assume, for simplicity, 
that $\sin^2{2\theta}\sim1$ and $(E_{\tilde\gamma}-E_\gamma)
\sim m_{\tilde\gamma}c^2\gg\hbar\Delta t^{-1}_{coll}$, so that the
fast oscillating cosine term vanishes upon time and ensemble averaging.
However, the mixing angle and oscillation time contributions may 
be trivially recovered when needed.

First, let us consider
${\gamma\gamma\rightleftarrows\tilde\gamma\tilde\gamma}$ 
process. In equilibrium, according to (\ref{condition2}), 
${\langle\sigma v\rangle}_{\gamma\gamma}\n^2_\gamma =
{\langle\sigma v\rangle}_{\tilde\gamma\tilde\gamma}\n^2_{\tilde\gamma}$,
where $\sigma$ is the effective cross-section of a process. The 
characteristic times are $\tau_{\gamma\to\tilde\gamma}^{-1}\sim
{\langle\sigma v\rangle}_{\gamma\gamma}\n_\gamma$ and
$\tau_{\tilde\gamma\to\gamma}^{-1}\sim
{\langle\sigma v\rangle}_{\tilde\gamma\tilde\gamma}\n_{\tilde\gamma}$,
and $3H$ may be replaced [due to condition (\ref{condition1})] with
the Compton scattering inverse rate 
$\tau^{-1}_c\sim{\langle\sigma_c v_e^{th}\rangle}\n_e$, 
where $\sigma_c$ is the Compton (Thompson) cross-section
and $v_e^{th}$ is the electron thermal velocity.
Then, from the definition of $\tau_\gtg$ and (\ref{condition2}) and
using $p_f\simeq\n_{\tilde\gamma}/\n_\gamma$, we express the effective
cross-sections as follows:
\begin{equation}
\frac{{\langle\sigma v\rangle}_{\tilde\gamma\tilde\gamma}}%
{\langle\sigma_c v_e^{th}\rangle}
\gtrsim\frac{1}{p_f^2}\frac{\n_e}{\n_\gamma}\sim10^7\ , \quad
\frac{{\langle\sigma v\rangle}_{\gamma\gamma}}%
{\langle\sigma_c v_e^{th}\rangle}
\gtrsim\frac{\n_e}{\n_\gamma}\sim10^{-9}\ .
\end{equation}
Hence, the $\tilde\gamma\tilde\gamma$ cross-section is too large
compared to that expected \cite{gg} from particle physics experiments.
[Such a cross-section corresponds to the photino mass (using \cite{gg}) 
$m_{\tilde\gamma}\lesssim0.5\textrm{~eV}/c^2$, which contradicts to 
(\ref{bound-m}).] Thus, such a process is (probably) ruled out. 
Second, the same is true for 
${\gamma\tilde x\rightleftarrows\tilde\gamma x}$ process. Indeed, 
the cross-sections (estimated for electrons, $x\equiv e$) are
\begin{equation}
\frac{{\langle\sigma v\rangle}_{\tilde\gamma e}}%
{\langle\sigma_c v_e^{th}\rangle}
\gtrsim\frac{1}{p_f}\sim10^8\ , \quad
\frac{{\langle\sigma v\rangle}_{\gamma\tilde e}}%
{\langle\sigma_c v_e^{th}\rangle}
\gtrsim\frac{\n_e}{\n_{\tilde e}}\ ,
\end{equation}
which are also too large. Third, the situation is better for 
${\tilde\gamma\rightleftarrows\tilde a\gamma}$ process. Denoting
$\Gamma_{\tilde\gamma}^{decay}$ to be the $\tilde\gamma$-decay rate,
we obtain:
\begin{equation}
\Gamma_{\tilde\gamma}^{decay}
\gtrsim\frac{3H_c}{p_f}\sim10^{-2}~\textrm{sec}^{-1}\ , \quad
\frac{{\langle\sigma v\rangle}_{\tilde a\gamma}}%
{\langle\sigma_c v_e^{th}\rangle}
\gtrsim\frac{\n_e}{\n_{\tilde a}}\ ,
\end{equation}
where $H_c=H(z_c)$. Since the axino is a 
possible non-barionic dark matter candidate, the ratio 
$\n_e/\n_{\tilde a}$ may be quite small, too. Estimated from the
$\tilde\gamma$-decay rate \cite{ag}, the photino mass is
$m_{\tilde\gamma}\gtrsim3\textrm{~MeV}/c^2$
(for the Peccei-Quinn symmetry breaking scale $V_{PQ}=10^9$~GeV). 
Since $m_{\tilde\gamma}c^2\gg kT_c$,
fine-tuning $m_{\tilde\gamma}\simeq m_{\tilde a}$ is required; 
otherwise the inverse decay ${\tilde a\gamma\to\tilde\gamma}$
is strongly suppressed. Assuming 
$m_{\tilde\gamma}\sim m_{\tilde a}\sim3\textrm{~MeV}/c^2$, 
we may estimate the
cosmological abundance of axinos (for the $\Omega_{tot}=1$ Universe) to be 
$\n_{\tilde a}/\n_e\sim\Omega_0^{-1}m_{barion}/m_{\tilde a}\lesssim10^5$,
which yields a reasonable $\tilde a\gamma$ cross-section.

It is important to note that the restrictions imposed by condition
(\ref{condition2}) may, however, be significantly relaxed by
admitting (i) multi-channel $\ggt$ inter-conversions (e.g.,
${\gamma\to\tilde\gamma}$ and ${\tilde\gamma\to\gamma}$ conversions
may proceed through different processes), (ii) super-symmetry theories
beyond the standard model, or (iii) theories with R-parity violation.
Detailed study of these issues is, obviously, beyond the scope of this
Letter.

To conclude, we have shown that, under appropriate conditions, the
 $\ggt$ inter-conversions may result in a detectable distortion of the CMBR
blackbody energy spectrum. Since the `bare' 
${\gamma\rightleftarrows\tilde\gamma}$ process is forbidden
by angular momentum and R-parity conservation, kinetics of super-symmetric
particle--matter interactions plays an important role. Thus, the amplitude
of the CMBR spectrum distortion is related to photino properties, e.g., 
its mass, $m_{\tilde\gamma}$. 
Recent data from the COBE-mission are used to evaluate the 
distortion parameter $\sqrt{p_f}$ and photino mass.

Author is grateful to K. Griest for valuable and interesting discussions and to
P.H. Diamond and A. Wolfe for their interest in this work.
This work was supported by DoE grant DE-FG03-88ER53275.


\begin{references}
\bibitem{c2} 
There is a typing error in equation (11) in \cite{ambiguous}. 
It should read $x_j=\left[n_j(\rho+p_f)-\delta\right]/(2\rho+\delta)$.
\end{references}
\end{document}